# SPIN STUDIES OF NUCLEONS IN A STATISTICAL MODEL

J.P. Singh   and   Alka Upadhyay

Physics Department,   Faculty of   Science
M.S.  University of  Baroda, Vadodara-390002, India

## Abstract

We decompose  various quark–gluon  Fock  states of a  nucleon in a set of states in which each of the three-quark core and the rest of the stuff, termed as a sea, appears with definite  spin  and  color quantum   number,  their  weight  being  determined,  statistically,  from  their  multiplicities.  The expansion coefficients in the  quark-gluon Fock states have been taken from a  recently proposed statistical model. We have also considered  two modifications of this model with a view to reduce the contributions of the sea components  with  higher  multiplicities. With certain  approximations, we have calculated the  quark contributions to the spin  of the nucleon, the ratio of the magnetic moments of  nucleons, their weak decay constant, and the ratio of SU(3) reduced matrix elements for the axial current.  A  reasonably  close  agreement  with  the  corresponding  experimental  values  have  been obtained  in all the three cases.

## I. INTRODUCTION

The composition of nucleons, in terms of fundamental quarks and gluons degrees of freedom have been modeled variously to account for its observed properties. It is important to calculate as many nucleonic parameters as possible in these models to check their merits and their domains of validity. The naive valence picture of nucleon structure may be regarded as a first order approximation to the real system[1]. Models with one constituent gluon [2] and with one quark- antiquark $q\bar{q}$ pair [3-5], in addition to the three valence quarks, are capable of giving better account of nucleonic properties. In another class of models, it is assumed that nucleons consist of valence quarks surrounded by a "sea"which, in general, contains gluons and virtual quark-antiquark pairs, and is characterized by its total quantum number consistent with the quantum number of nucleons[6,7].

In the chiral quark model of Manohar and Georgi [8], QCD quarks propagate in the nontrivial QCD vaccum having $q\bar{q}$ condensates and this leads to the generation of extra mass to the quarks. As a consequence of this spontaneous chiral symmetry breaking, massless pseudoscalar bound $q\bar{q}$ Goldstone bosons are generated, and this leads to the nontrivial sea structure of the nucleon. In the instanton model [9], the quark-antiquark sea in a nucleon results from a scattering of a valance quark off a nonperturbative vaccum fluctuation of the gluon field, instanton. In the instanton induced interaction described by 't Hooft effective lagrangian, the flavor of the produced quark-antiquark is different from the flavor of the initial valance quarks, and there is a specific correlation between the sea quark helicity and the valance quark helicity. In the chiral-quark soliton model [10], the large $N_c$ model of QCD becomes an effective theory of mesons with the baryons appearing as solitons. Quarks are described by single particle wave functions which are solutions of the Dirac equation in the field of the background pions. In the statistical approach, the nucleon is treated as a collection of massless quarks, antiquarks and gluons in thermal equilibrium within a finite size volume[11]. The momentum distributions for quarks and antiquarks follow a Fermi-Dirac distribution function characterized by a common temperature and a chemical potential which depends on the flavor and helicity of the quarks.

Recently, a new statistical model has been proposed in which a nucleon is taken as an ensemble of quark-gluon Fock states [12,13]. In this model, using the principle of balance that every Fock state should be balanced with all of the nearby Fock states[13], or using the principle of detailed balance that any two nearby Fock states should be balanced with each other[12], the probability of finding every Fock state of the proton accounting upto $\approx 98\%$ of the total Fock state has been obtained. It has been shown that the model gives an excellent description of the light flavor sea asymmetry (i.e, $\bar{u} \neq \bar{d}$ ) without any parameter [12,13]. In this article, we have used this model to calculate the light quark spin content of nucleons, the ratio of their magnetic moments, the semileptonic decay constant of neutron, and the ratio of SU(3) reduced matrix elements for the axial current.

## II. SEA AND ITS STRUCTURE

In Ref.[12,13], treating the proton as an ensemble of quark-gluon Fock states, the proton state has been expanded in a complete set of such states as

$$| p \rangle = \sum_{ijk} C_{ijk} | uud,i,j,k \rangle ,$$

where i is the number of $\bar{u}u$ pairs, j is the number of $\bar{d}d$ pairs, and k is the number of gluons.

The probability to find a proton in the Fock state $|\text{uud,i,j,k}\rangle$ is

$$\rho_{ijk} = |C_{ijk}|^2,$$

where $\rho_{ijk}$ satisfies the normalization condition,

$$\sum_{ijk} \rho_{ijk} = 1.$$

Then using the detailed balance principle or balance principle, and with subprocesses $q \Leftrightarrow q\,g$, $g \Leftrightarrow \bar{q}q$ and $g \Leftrightarrow gg$ considered, all $\rho_{ijk}$ have been calculated explicitly. Interestingly, the model predicts an asymmetry in the sea flavor of $\bar{u}$ and $\bar{d}$ as $\bar{d} - \bar{u} \sim 0.124$ in surprising agreement with the experimental data $0.118 \pm 0.012$. These quarks and gluons have to be understood as "intrinsic" partons of the proton as opposed to the "extrinsic" partons generated from the QCD hard bremsstrahlung and gluon splitting as a part of the lepton nucleon scattering interaction[14]. The $\bar{q}q$ pairs and gluons, which are multiconnected non-perterbatively to the valence quarks, will collectively be referred to as the sea. Since the proton should be colorless and a $q^3$ state can be in color state $1_c$, $8_c$ and $10_c$, the sea should also be in the corresponding color state to form a color singlet proton. Furthermore, if the sea is in an S-wave state relative to the $q^3$ core, conservation of angular momentum restricts that the spin of the sea can only be 0,1or 2 to give a spin-1/2 proton. The case of the sea with one $\bar{q}q$ pair, where the sea or at least one of the quarks is needed to be in a relative P-wave to meet the positive parity requirement of the proton, will be treated separately. We take the probabilities of finding various quark-gluon Fock states in a proton from Ref.[13], and assume that the quarks and the gluons can be treated nonrelativistically for our problem, and also that, in general, these are in S-wave motion . The case of a neutron will be treated in an analogous way using isospin symmetry.

Nonrelativistic treatments of quarks in nucleon models are well known [1,4-6]. There are phenomenological evidences that gluons also behave as massive particles with mass $\geq 0.5\text{GeV}$[15].

There is a firm evidence from lattice calculation also that gluons behave as massive particles at low momenta ($\leq$4GeV)[16]. It has been shown in Ref [5] that the sum of the relativistic quark spin and orbital angular momentum (derived from QCD Lagrangian ) is equal to the sum of the non relativistic quark spin and orbital angular momentum,

$$\vec{S}_q + \vec{L}_q = \vec{S}^{\,NR}_q + \vec{L}^{\,NR}_q$$

Furthermore, it has been shown that on truncating the Fock space to contain only $|q^3\rangle$ and $|q^3 \bar{q}q\rangle$ component, the quark orbital angular momentum contribution comes out to be negligible or small [5]. This contribution should decrease on inclusion of Fock states with more "intrinsic" partons, since then each parton will have a lesser linear momentum share, and hence, smaller orbital angular momentum too.

Following Ref.[6] we write the possible combination of $q^3$ and sea wave function, which can give a spin ½ flavor octet, color singlet state as

$\Phi_1^{(1/2)}H_0G_1$, $\Phi_8^{(1/2)}H_0G_8$, $\Phi_{10}^{(1/2)}H_0G_{\bar{10}}$, $\Phi_1^{(1/2)}H_1G_1$, $\Phi_8^{(1/2)}H_1G_8$, $\Phi_{10}^{(1/2)}H_1G_{\bar{10}}$ and

$\Phi_8^{(3/2)}H_1G_8$, $\Phi_8^{(3/2)}H_2G_8$.

In the above $\Phi^{(1/2,3/2)}_{1,8,10}$ is the $q^3$ wave function in obvious notation, while $H_{0,1,2}$ and $G_{1,8,\bar{10}}$ denote spin and color sea wave functions respectively [6] which satisfy

$$\langle H_i \mid H_j \rangle = \delta_{ij}, \quad \langle G_k \mid G_l \rangle = \delta_{kl}$$

The total flavor-spin-color wave function of a spin up proton which consists of three valence quarks and sea component can be written as [6]:

$| \Phi_{1/2}^{\uparrow} \rangle = (1/N) [\Phi_1^{(1/2\uparrow)}H_0G_1 + a_8 \Phi_8^{(1/2\uparrow)}H_0G_8 + a_{10} \Phi_{10}^{(1/2\uparrow)}H_0G_{\bar{10}} + b_1( \Phi_1^{(1/2)} \otimes H_1)^{\uparrow}G_1 +$

$b_8(\Phi_8^{(1/2)} \otimes H_1)^{\uparrow}G_8 + b_{10} (\Phi_{10}^{(1/2)} \otimes H_1)^{\uparrow}G_{\bar{10}} + c_8 (\Phi_8^{(3/2)} \otimes H_1)^{\uparrow}G_8 + d_8(\Phi_8^{(3/2)} \otimes H_2)^{\uparrow}G_8]$

$$......(1)$$

where $N^2 = 1 + a_8^2 + a_{10}^2 + b_1^2 + b_8^2 + b_{10}^2 + c_8^2 + d_8^2$,

and $(\Phi_1^{(1/2)} \otimes H_1)^\uparrow$, etc. have to be written properly with appropriate CG coefficients and by taking into account the symmetry property of the component wave function [6] . Furthermore, we will use an approximation in which quarks in the $q^3$ core will not be antisymmetrised with the identical quarks appearing in the sea. Use of different labels for valance and sea quarks has been justified with the assumption that the valance and the sea quarks have very different momentum distributions, with the valance quarks being "hard" and the sea quarks "soft", and that the overlap region between the two momentum distributions is negligible [17]. Consequently, this classification can work where one is concerned with matrix elements having zero momentum transfer and only require that the overlap region between valance and sea quark momentum distribution be negligibly small. Nevertheless, we will use this separation for the problem of quark contribution to the nucleon spin as well.

Next, we decompose each one of the Fock states $|uud,i,j,k\rangle$ in terms of the above set of states following a statistical approach .

(i) Consider the decomposition of a state $|uud,0,0,2\rangle$ or $|gg\rangle$ sea (two gluons in the sea ).

Spin : uud : $1/2 \otimes 1/2 \otimes 1/2 = 2(1/2) \oplus 3/2$,

    gg : $1 \otimes 1 = 0 \oplus 1 \oplus 2$.

Color : uud : $3 \otimes 3 \otimes 3 = 1 \oplus 8 \oplus 8 \oplus 10$,

    gg : $8 \otimes 8 = 1_s \oplus 8_s \oplus 8_a \oplus 10_a \oplus \overline{10}_a \oplus 27_s$ .

The subscripts s and *a* denote symmetry and asymmetry respectively under the exchange of two identical bosons (gluons above). Call $\rho_{j1\,j2}$ as the probability that the $q^3$ core and gg sea are in angular momentum states $j_1$ and $j_2$ respectively, and they finally add to give total angular momentum 1/2. Let us compare such probabilities.

$$\rho_{1/2\,0}\,/\,\rho_{1/2\,1} = \frac{(4/8).(1/9).1}{(4/8).(3/9).(2/6)} = 1,$$

$$\rho_{1/2\,0}\,/\,\rho_{3/2\,2} = \frac{(4/8).(1/9).1}{(4/8).(5/9).(2/20)} = 2,$$

$$\rho_{3/2\,1}\,/\,\rho_{3/2\,2} = \frac{(4/8).(3/9).(2/12)}{(4/8).(5/9).(2/20)} = 1,$$

$$\rho_{1/2\,1}\,/\,\rho_{3/2\,1} = \frac{(4/8).(3/9).(2/6)}{(4/8).(3/9).(2/12)} = 2.$$

The first factor in the numerator or denominator in the r.h.s is the relative probability for the core quarks to have spin $j_1$, the second factor is the same for the two gluons to have spin $j_2$, and finally the third one is the same for $j_1$ and $j_2$ to have resultant 1/2. In future, we will omit the factor which is common in the numerator and the denominator.

Similarly we can compare the probabilities for the $q^3$ core and gg to be in different color substates which finally give a color singlet proton. In obvious notations:

$$\rho_{1\,1}\,/\,\rho_{8\,8s} = \frac{(1/27).(1/64).1}{(16/27).(8/64).(1/64)} = 1/2 = \rho_{1\,1}/\rho_{8\,8a},$$

$$\rho_{1\,1}\,/\,\rho_{10\,\overline{10}} = \frac{(1/27).(1/64).1}{(10/27).(10/64).(1/100)} = 1.$$

The product of probabilities in spin and color spaces can be written in terms of one common parameter c as

$$\rho_{1/2\,0}\,[\rho_{1\,1},\,\rho_{8\,8s}] = 2c\,(1,2),$$

$$\rho_{1/2\,1}\,[\,\rho_{8\,8a},\,\rho_{10\,\overline{10}}] = 2c\,(2,1),$$

$$\rho_{3/2\,1}\,[\rho_{8\,8a}] = 2c\,, \quad \rho_{3/2\,2}\,[\rho_{8\,8s}] = 2c.$$

There is no contribution to $H_0 G\bar{_{10}}$ and $H_1 G_1$ sea from two gluon states because $H_0$ and $G_1$ are symmetric whereas $H_1$ and $G\bar{_{10}}$ are antisymmetric under exchange of the two gluons making these product wave functions antisymmetric and hence unacceptable for a bosonic system.

The sum of all these probabilities is taken from Ref.[13] and this determines the unknown parameter c :

$\rho_{uud\,gg}$ = 0.081887,    c = 0.005118.

Similar decomposition will hold good for $\bar{qq}\,\bar{qq}$ sea also.

(ii) For decomposition of $\left| g, \bar{qq} \right\rangle$ and $\left| \overline{uu}\,\overline{dd} \right\rangle$ sea, symmetry consideration is not needed. Here we have assumed that $\bar{qq}$ carries the quantum numbers of a gluon due to the subprocesses $g \Leftrightarrow \bar{qq}$. This gives the relative probability density in color space as $\rho_{1\,1}/\rho_{8\,8}$ =1/4.

The ratio $\rho_{1\,1}/ \rho_{10\,\overline{10}}$ and the relative densities in spin space remain the same as in (i). The products of densities in spin and color spaces come out as

$\rho_{1/2\,\,0}\,[\rho_{1\,1},\,\rho_{8\,8},\,\rho_{10\,\overline{10}}]$ = 2c (1,4,1),

$\rho_{1/2\,1}\,[\rho_{11},\,\rho_{8\,8},\,\rho_{10\,\overline{10}}]$ = 2c (1,4,1),

$\rho_{3/2\,1}\,[\rho_{8\,8}]$ = 4c,    $\rho_{3/2\,2}\,[\rho_{8\,8}]$= 4c.

Equating the sum of the above partial probabilities to $\rho_{101}$, $\rho_{011}$ and $\rho_{110}$ from Ref.[13], we get the respective values of c as

c = 0.001718, 0.002585, 0.000916.

(iii) $\left| g\,g\,\bar{qq} \right\rangle$, $\left| \bar{qq}\,\bar{qq}\,g \right\rangle$ sea : First we take the product of two spin 1 states and two color octet states as in (i). These are further multiplied with spin 1 and color octet state respectively. The new results needed are

Spin :     $1 \otimes 2 = 1 \oplus 2 \oplus 3$,

Color:     $10 \otimes 8 = 8 \oplus 10 \oplus 27 \oplus 35$,

$$27 \otimes 8 = 8 \oplus 10 \oplus \overline{10} \oplus 2(27) \oplus 35 \oplus 35 \oplus 64 \; .$$

Using the subscript s and $a$ for symmetry and asymmetry under the exchange of first two bosons, the relative probability densities in spin space are:

$$\rho_{1/2\,0a} / \rho_{1/2\,1a} = \frac{(1/27).1}{(3/27).(2/6)} = 1, \quad \rho_{1/2\,0a} / \rho_{1/2\,1s} = \frac{(1/27).1}{(6/27).(4/12)} = \frac{1}{2},$$

$$\rho_{1/2\,1a} / \rho_{3/2\,1a} = \frac{(3/27).(1/3)}{(3/27).(1/6)} = 2 = \rho_{1/2\,1s} / \rho_{3/2\,1s},$$

$$\rho_{3/21a} / \rho_{3/2\,2a} = \frac{(3/27).(2/12)}{(5/27).(2/20)} = 1, \quad \rho_{3/2\,1s} / \rho_{3/2\,2s} = \frac{(6/27).(4/24)}{(5/27).(2/20)} = 2 \; .$$

The ratio of the probability densities in color space are:

$$\rho_{1\,1s} / \rho_{8\,8s} = \frac{(1/27).(1/512).1}{(16/27).(32/512).(1/64)} = 1/8,$$

$$\rho_{1\,1s} / \rho_{10\,\overline{10}\,s} = \frac{(1/27).(1/512).1}{(10/27).(20/512).(1/100)} = \frac{1}{2} = \rho_{1\,1a} / \rho_{10\,\overline{10}\,a} \; ,$$

$$\rho_{1\,1a} / \rho_{8\,8a} = \frac{(1/27).(1/512).1}{(16/27).(32/512).(1/64)} = 1/8 \; .$$

The combined probabilities in spin and color space can be written in terms of a common factor c as

$$\rho_{1/2\,0a} \left[ \rho_{1\,1a}, \rho_{8\,8a}, \rho_{10\,\overline{10}\,a} \right] = 2c(1,8,2) \; ,$$

$$\rho_{1/2\,1a} \left[ \rho_{1\,1a}, \rho_{8\,8a}, \rho_{10\,\overline{10}\,a} \right] = 2c(1,8,2) \; ,$$

$$\rho_{1/2\,1s} \left[ \rho_{1\,1s}, \rho_{8\,8s}, \rho_{10\,\overline{10}\,s} \right] = 4c(1,8,2) \; ,$$

$$\rho_{3/2\,1a} \left[ \rho_{8\,8a} \right] = 8c, \quad \rho_{3/2\,1s} \left[ \rho_{8\,8s} \right] = 16c, \quad \rho_{3/2\,2a} \left[ \rho_{8\,8a} \right] = 8c.$$

Summing all the partial probabilities and equating it to the probabilities $\rho_{102}, \rho_{012}, \rho_{201}$ and $\rho_{021}$ from Ref.[13] we get

c = 0.000254, 0.000379, 0.000037, 0.000127.

(iv) $| \overline{uu} \, \overline{dd} \, g \rangle$ sea: Here, there is no symmetry requirement. Ratios of probability densities are

$\rho_{1/2\ 0}\ /\ \rho_{1/2\ 1} = 1/3, \quad \rho_{1/2\ 0}/\rho_{3/2\ 2} = 1, \quad \rho_{1/2\ 1}\ /\ \rho_{3/2\ 1} = 2, \quad \rho_{3/2\ 1}\ /\ \rho_{3/2\ 2} = 3/2.$

in spin space, and

$\rho_{1\ 1}\ /\ \rho_{8\ 8} = 1/8, \quad \rho_{1\ 1}\ /\ \rho_{10\ \overline{10}} = 1/2$

in color space. Their products can be written as

$\rho_{1/2\ 0}\ [\rho_{11}, \rho_{8\ 8}, \rho_{10\ \overline{10}}] = c(1,8,2)\ , \quad \rho_{1/2\ 1}[\ \rho_{11}, \rho_{8\ 8}, \rho_{10\ \overline{10}}] = 3c(1,8,2),$

$\rho_{3/2\ 1}\ [\ \rho_{8\ 8}] = 12c, \quad \rho_{3/2\ 2}\ [\rho_{8\ 8}] = 8c\ .$

Equating the sum of above product probabilities to the value given for $\rho_{111}$ in Ref.[13], we get

c = 0.000478.

(v) $|\,ggg\rangle$ sea : The wave function for this sea should be completely symmetric under the exchange of any two gluons . Among the product spin function, the total spin S= 0 is completely antisymmetric and one S=1 is completely symmetric . Among the product color functions, there is one color singlet state and one color octet state which are completely antisymmetric; and there is one color singlet state and one color octet state which are completely symmetric. This gives

$\rho_{1/2\ 0}\ /\rho_{1/2\ 1} = 1, \quad \rho_{1/2\ 1}/\rho_{3/2\ 1} = 2,$

$\rho_{1\ 1a,s}\ /\rho_{8\ 8} = 1/2,$

$\rho_{1/2\ 0a}\ [\rho_{1\ 1a}, \rho_{8\ 8a}] = c(1,\ 2), \quad \rho_{1/2\ 1s}[\ \rho_{11s}, \rho_{8\ 8s,}] = c(1,2),$

$\rho_{3/2\ 1s}\ [\ \rho_{8\ 8s}] = c.$

Equating the sum of above product probabilities to $\rho_{003}$ from Ref.[13], we get

c = 0.005343.

A confined gluon in the sea may be divided into TE (transverse electric ) modes with $J^{pc} = 1^{+-}$ and the TM (transverse magnetic) modes with $J^{pc} = 1^{--}$. The Fock states with a single gluon in the sea may be considered to be consisting of a TE gluon [18]. Clearly, a gluon in the sea will contribute only to the $H_1G_8$ component of the sea . From this decomposition we get the following numbers for the coefficients in the expansion (1) of the proton state:

$a_8{}^2$=0.5043,     $a_{10}{}^2$= 0.0892,   $b_1{}^2$=0.1037,   $b_8{}^2$ =1.8133,    $b_{10}{}^2$ = 0.222,

$c_8{}^2$ = 0.90668,   $d_8{}^2$ = 0.26304  and    $N^2$ = 4.9024.

However, the treatment of a $\overline{q}q$ pair in the sea requires special attention, since as stated earlier, to keep the parity of the system positive, one or a group of the five particles is required to be in a P-wave state. This requires detailed knowledge of spatial wave function. To get the contribution of this particular Fock space, we have borrowed the result from Ref.[5] and scaled it to give the same probability which we are using, as given in Ref.[13]. Unlike our treatment, the total wave function in [5] has been properly antisymmetrised. All the above states taken together constitute ≈ 86% of the total Fock space .

We have tabulated the values of α and β, as defined in Ref [6], in Table I (model C ). These can be used to calculate various physical quantities as done in Ref.[6], where the sea plays a role of "passive" background and the relevant operators act only on the three-quark core. When the operator $\sum_i e_i{}^2\sigma_z^i$ acts on the sea minus the single $\overline{q}q$ component, i.e. when the sea plays the "active" role, the result has been denoted by $\Delta I_1^p$ and $\Delta I_1^n$ for the proton and neutron respectively .There is no such contribution to the magnetic moments due to the active sea , since the $\overline{q}q$ pairs carry the quantum numbers of the parent gluons. The total contribution to the nucleon spin from the spins of the quarks, denoted by $I_1^p$ and $I_1^n$, has been displayed in table II and compared with the revised EMC result [19]. We should note that EMC value is for $Q^2$ ≈10 GeV$^2$ which can be very different from the low energy result we have obtained for $I_1^p$ and $I_1^n$. To estimate ($g_A/g_V$), we use Bjorken sum rule written upto $O(\alpha_s^3/\pi^3)$ [20]. We have considered three values of $\alpha_s$ from the recent literature. Authors of Ref.[21] have used $\alpha_s$ (1GeV$^2$ ) ≈ 0.05 for the same purpose as ours. Particle Data Group[22] average value is $\alpha_s(m_c)$=0.357,  which we modify as $\alpha_s$(1GeV$^2$)=0.375 for our use. Authors of Ref.[23] use $\alpha_s$(0)=0.35 (to fit the bound states in QCD). The values of ($g_A/g_V$) obtained for each one of these values have been displayed in Table II . The F/D value has been obtained from α and β as per the prescription given in Ref.[6 ].

In order to check the stability of our results against some plausible changes in some physical parameters, we consider two modifications of the above model. It appears reasonable to assume that in determining low energy hadronic quantities, the long range and confining forces, in addition to the statistical consideration, will have a role to play. Based upon this point of view, we introduce the following two models:

**A. Sea with pseudoscalars**.

In the statistical formulation of Ref.[12,13], a quark-antiquark pair is created from a gluon splitting: g $\Leftrightarrow$ $\bar{q}q$ . This pair, naturally, carries the quantum numbers of the parent gluon. However, this is not an energetically favorable situation even within the hadronic boundary [24] ; the pair on exchange of a soft gluon with the rest of the system, and also possibly on a spin flip, will evolve to a colorless pseudoscalar form, called internal Goldstone boson [24-26]. We will assume that all the $\bar{q}q$ pairs are in one or the other pseudoscalar form practically for whole of their lifetimes giving no contribution to the spin or the color charge of the proton . In case of |gg $\bar{q}q$ ⟩ state, in order to compensate the odd parity of the $\bar{q}q$ pair, one of the gluons will be assumed to be in TE mode while the other in TM mode. It gives the following contribution to the expansion coefficients in (1) of the proton state:

$a_8^2 = 0.221434$, $\quad a_{10}^2 = 0.0216048$, $\quad b_1^2 = 0.0424686$, $\quad b_8^2 = 1.254083$, $\quad b_{10}^2 = 0.06825$,

$c_8^2 = 0.6270414$, $d_8^2 = 0.0898495$.

This sea will not "actively" contribute to the spins or the magnetic moments of the nucleons. With this sea, the results of the spin distribution of nucleons come closer to the data as is evident from

Table II. There is hardly any change in the values of the ratios $\mu_p/\mu_n$ and F/D from the previous case. Matching the values of $g_A/g_V$ with the experimental numbers favors the smaller values of $\alpha_s$.

## B. Sea with suppressed higher multiplicity states

We propose a second modification of the model in which the contribution to the states with higher multiplicities is suppressed. Within the hadronic boundary, pseudoscalar exchange have been found to dominate over vector exchange and even gluon exchanges [5,24-26]. Although we are not using any dynamical model, we tend to believe that the states with larger number of gluons (having corresponding smaller probabilities) approximate the ones with saturated gluons for which color neutrality is achieved over a certain scale, which is called "saturation scale" [27,28]. In Landshoff- Nachtmann model, quark–quark and hadron–hadron scatterings are assumed to arise due to exchanges of two non-perturbative gluons having vacuum quantum numbers[29]. It is believed that pomeron and odderon exchanges are associated with the exchanges of a family of glueballs which are colorless but of different spins [29]. It is reasonable to assume that when a set of "intrinsic" gluons exist in a nucleon, they would prefer to be in a similar state.

Even within the hadronic boundary, Goldston boson exchange (GBE) model successfully describes diverse phenomenon [24-26 ]. In color space, singlets are unique due to confinement , but even there the color octet exchange models, and not any higher color states exchange model, have been successfully used [30] . Larger is color multiciplity of a group of particles (here the sea), larger will be the probability of its interaction with the rest of the particles (the core) and smaller will be its probability of survival. Authors of Ref.[6] have, on phenomenological ground, proposed a set of parameters in which states with higher multiplicities occur with lower probabilities.

In view of these phenomenological evidences, it appears reasonable to propose that higher multiciplity states are suppressed. We parameterize this suppression in a simple way by assuming that probability of a system to be in a spin and color state is inversely proportional to the multiciplity (both in spin and color spaces) of the state. This probability factor is additional to the previously incorporated factors in the probabilities. With this new input, we decompose Fock states as follows.

(i) $| \, gg \rangle$, $| \, \overline{qq} \, \overline{qq} \, \rangle$ sea :

$\rho_{1/2 \, 0s} [\rho_{11s}, \rho_{8 \, 8s}, ] = 2d(1, 1/32)$,

$\rho_{1/2 \, 1a} [\rho_{8 \, 8a}, \rho_{10 \, \overline{10} \, a}] = 2d(1/96, 1/300)$,

$\rho_{3/2 \, 1a} [\rho_{8 \, 8a}] = d/192$,    $\rho_{3/2 \, 2s} [\rho_{8 \, 8s}] = d/320$.

Equating the sum of the above product probabilities to $\rho_{uudgg}$, we get

d=0.03903

Similarly for $| \, \overline{uu} \, \overline{uu} \, \rangle$ sea, d = 0.00345 and for $| \overline{dd} \overline{dd} \, \rangle$ sea, d = 0.00694.

(ii) $| \, g \overline{qq} \, \rangle$, $| \, \overline{uu} \, \overline{dd} \, \rangle$ sea :

$\rho_{1/2 \, 0} [\rho_{11}, \rho_{8 \, 8}, \rho_{10 \, \overline{10}}] = 2d(1, 1/16, 1/100)$,

$\rho_{1/2 \, 1} [\rho_{11}, \rho_{8 \, 8}, \rho_{10 \, \overline{10}}] = 2d(1/3, 1/48, 1/300)$,

$\rho_{3/2 \, 1} [\rho_{8 \, 8}] = d/96$,    $\rho_{3/2 \, 2} [\rho_{8 \, 8}] = d/160$.

This gives d = 0.01912 for $| g \overline{uu} \rangle$ sea, d = 0.02876 for $| \overline{dd} \, g \rangle$ sea, and d = 0.010898 for $| \overline{uu} \, \overline{dd} \, \rangle$ sea.

(iii) $| \, gg \overline{qq} \, \rangle$, $| \, \overline{qq} \, \overline{qq} \, g \rangle$ sea :

$\rho_{1/2 \, 0a} [\rho_{11a}, \rho_{8 \, 8a}, \rho_{10 \, \overline{10} \, a}] = d(1, 1/8, 1/50)$,

$\rho_{1/2 \, 1s} [\rho_{11s}, \rho_{8 \, 8s}, \rho_{10 \, \overline{10} \, s}] = d(1, 1/8, 1/50)$,

$\rho_{3/2 \, 1a} [\rho_{8 \, 8a}] = d/32$,    $\rho_{3/2 \, 1s} [\rho_{8 \, 8s}] = d/32$,    $\rho_{3/2 \, 2a} [\rho_{8 \, 8a}] = d/160$.

This gives d=0.00328 for $|\overline{uu}\ \overline{uu}\ g\rangle$ sea, d=0.00655 for $|\overline{dd}\overline{dd}\ g\rangle$ sea , d=0.01952

for $|gg\overline{dd}\ \rangle$ sea and d=0.013069 for $|gg\overline{uu}\rangle$ sea.

(iv) $|\ g\overline{uu}\ \overline{dd}\ \rangle$ sea :

$\rho_{1/2\ 0}\ [\rho_{1\ 1}, \rho_{8\ 8}, \rho_{10\ \overline{10}}] = d(1/2, 1/16, 1/100)$,

$\rho_{1/2\ 1}\ [\ \rho_{1\ 1}, \rho_{8\ 8}, \ _{10\ \overline{10}}] = d(1/2, 1/16, 1/100)$,

$\rho_{3/2\ 1}\ [\ \rho_{8\ 8}\ ] = d/64$,    $\rho_{3/2\ 2}\ [\rho_{8\ 8}\ ] = d/160$.

This gives d = 0.026197.

(v)  $|ggg\rangle$ sea:

$\rho_{1/2\ 0a}\ [\rho_{1\ 1a}, \rho_{8\ 8a}] = d(1, 1/32)$,    $\rho_{1/2\ 1s}\ [\ \rho_{1\ 1s}, \rho_{8\ 8s}] = d/3(1, 1/32)$,    $\rho_{3/2\ 1s}\ [\rho_{8\ 8s}] = d/384$.

This gives d = 0.023269.

We would like to point out that there is nothing special about the use of the inverse of the multiplicity for suppression of higher multiplicity states. One could have fine tuned the power of the multiplicity to fit the data in a better way. It is only a possible way to suppress the contribution of states with higher multiplicities within the nucleon sea, which might be originally due to some dynamics. In the above calculation we have also included the (active) contribution of sea quarks.

III. SUMMARY AND CONCLUSION .

The statistical approach advocated in Ref.[12,13] was successful in describing the large asymmetry between $\overline{u}$ and $\overline{d}$ quark distributions of the proton. We have extended that approach by decomposing various quark-gluon Fock states into states in which the three quark core and the rest of the stuff (called sea ) have definite spin and color quantum numbers, using the assumption of equal probability for each substate of such a state of the nucleon. We have further used the approximation in which a quark in the core is not antisymmytrised with an identical quark in the sea, and have treated quarks and gluons as nonrelativistic particles moving in S-wave (except for a single $\bar{q}q$ sea) motion. Also we have not taken into account any contribution of the s-quark and other heavy quarks, and we have covered only $\approx 86\%$ of the total Fock state. With these approximations we have calculated the quarks contribution to the spin of the nucleons, the ratio of the magnetic moments of the nucleons, their weak decay constant, and the ratio of SU(3) reduced matrix elements for the axial current. All of these quantities give integrated result of Bjorken variable, as was the case for the flavor asymmetry in the nucleon sea calculated in Ref.[12,13]. We have also considered two modifications of the above statistical approach with a view to reduce the contributions of the sea components with higher multiplicities, and have done the above calculations for those two cases as well.

Our results of calculation holds good for a typical hadronic energy scale~1 GeV$^2$ [13]. Experimental result for $I_1^p$ and $I_1^n$ apply for $Q^2 \approx 10$ GeV$^2$, and their values will increase when evolved to a lower energy scale. Hence, our calculated result for $I_1^p$ and $I_1^n$ may well be consistent with the data. Our result for the ratio of magnetic moments of nucleons is within few percent of the data. Weak decay constant has been calculated using Bjorken sum rule, written up to $O(\alpha_s^3/\pi^3)$. There is some controversy in the value of $\alpha_s$ at the low energy~1GeV we are working at, and we have chosen three typical values taken from recent literature. The results for the weak decay constant are scattered over~2% to~20% from the experimental value for different cases. However, we should keep in mind that the use of Bjorken sum rule is not expected to give an accuracy better than 10% [20]. For the F/D ratio the maximum difference from the experimental value we have got is for the D-model, where our result is the same as obtained in chiral quark model with SU(3) symmetry [25]. In summary, results obtained are fairly stable within these models, and reasonably close to the experimental values.

ACKNOWLEDGMENT

Authors gratefully acknowledge the financial support by Department of Science and Technology, New Delhi.

### RFERENCES

TBLE I : $\alpha$ and $\beta$ as defined in Ref.[6]: $\alpha_1$ and $\beta_1$ are the contributions from the sea excluding the single $\overline{qq}$ components; $\alpha_2$ and $\beta_2$ are the contribution from the single $\overline{qq}$ components of the sea. $\Delta I_1^p$ and $\Delta I_1^n$ are the contribution to $I_1^p$ and $I_1^n$ respectively when the operator $\sum_i e_i^2 \sigma_z^i$ acts on the sea excluding the single $\overline{qq}$ component. Model C is our first statistical model described in the text. In model P, $\overline{qq}$ pairs have been taken as colorless pseudoscalars, whereas model D is the one in which suppressed higher multiplicity states appear.

| Model Type | $\alpha_1$ | $\alpha_2$ | $\alpha = \alpha_1 + \alpha_2$ | $\beta_1$ | $\beta_2$ | $\beta = \beta_1 + \beta_2$ | $\Delta I_1^p$ | $\Delta I_1^n$ |
|---|---|---|---|---|---|---|---|---|
| Model C | 0.182069 | 0.041667 | 0.223736 | 0.054919 | 0.018633 | 0.073552 | 0.03080 | 0.04059 |
| Model P | 0.213558 | 0.041667 | 0.255225 | 0.065956 | 0.018633 | 0.084589 | 0.00000 | 0.00000 |
| Model D | 0.222278 | 0.041667 | 0.263945 | 0.052113 | 0.018633 | 0.070746 | 0.01510 | 0.01790 |

TABLE II : Comparison of our calculated results of various physical parameters with the experimental numbers.

| Model Type | $I_1^p$ | $I_1^n$ | $\mu_p / \mu_n$ | $g_A/g_V$ | | | F/D |
|---|---|---|---|---|---|---|---|
| | | | | $\alpha_s = 0.35$ | $\alpha_s = 0.375$ | $\alpha_s = 0.5$ | |
| Model C | 0.1677 | 0.0291 | -1.4050 | 1.01853 | 1.04538 | 1.24328 | 0.60330 |
| Model P | 0.1561 | -0.0139 | -1.4022 | 1.24907 | 1.28199 | 1.52468 | 0.60134 |
| Model D | 0.1792 | 0.0147 | -1.4765 | 1.20957 | 1.24145 | 1.47647 | 0.65100 |
| Expt. Value [Ref.] | 0.136 [19] | -0.030 [19] | -1.4600 [22] | 1.2670 [22] | | | 0.57500 [25] |